\documentclass[aps,prd,twocolumn,floatfix,preprintnumbers,altaffilletter,nofootinbib]{revtex4-1}
\usepackage{graphicx,url,amssymb,amsmath,longtable,rotating,color,units,wasysym,subfigure,epsfig,multirow,epstopdf,bm}
\usepackage[colorlinks,urlcolor=blue,citecolor=blue,linkcolor=blue]{hyperref}
\usepackage{cancel}
\usepackage{units}
\usepackage[T1]{fontenc}
\usepackage{xspace}
\usepackage{natbib}
\usepackage{hhline}
\usepackage[dvipsnames]{xcolor}
\usepackage[normalem]{ulem}

\newcommand{\textodds}{astrophysical odds\xspace}

\newcommand{\hyp}{\mathcal{H}}

\newcommand{\Hs}{\mathcal{S}}

\newcommand{\Hnots}{\cancel{\mathcal{S}}}

\newcommand{\Hg}{\mathcal{G}}
\newcommand{\Hnotg}{\cancel{\Hg}}
\newcommand{\Hn}{\mathcal{N}}

\newcommand{\odds}{\mathcal{O}}
\newcommand{\data}{d}
\newcommand{\alldata}{\bm{d}}

\newcommand{\pasa}{Publ. Astron. Soc. Aust.}
\newcommand{\apjs}{Astrophys. J., Suppl.}
\newcommand{\cqg}{Class. Quantum Gravity}
\newcommand{\prx}{Phys. Rev. X}
\newcommand{\mnras}{Mon. Notices Royal Astron. Soc.}

\begin{document}

\title{Gravitational wave detection without boot straps: a Bayesian approach}

\author{Gregory Ashton}
\email{greg.ashton@monash.edu}
\author{Eric Thrane}
\email{eric.thrane@monash.edu}
\author{Rory J. E. Smith}
\email{rory.smith@monash.edu}
\affiliation{Monash Centre for Astrophysics, School of Physics and Astronomy, Monash University, VIC 3800, Australia}
\affiliation{OzGrav: The ARC Centre of Excellence for Gravitational-Wave Discovery, Clayton, VIC 3800, Australia}
\date{\today}

\begin{abstract}

In order to separate astrophysical gravitational-wave signals from instrumental noise, which often contains transient non-Gaussian artifacts, astronomers have traditionally relied on bootstrap methods such as time slides. Bootstrap methods sample with replacement, comparing single-observatory data to construct a background distribution, which is used to assign a false-alarm probability to candidate signals. While bootstrap methods have played an important role establishing the first gravitational-wave detections, there are limitations. First, as the number of detections increases, it makes increasingly less sense to treat single-observatory data as bootstrap-estimated noise, when we know that the data are filled with astrophysical signals, some resolved, some unresolved. Second, it has been known for a decade that background estimation from time-slides eventually breaks down due to saturation effects, yielding incorrect estimates of significance. Third, the false alarm probability cannot be used to weight candidate significance, for example when performing population inference on a set of candidates. Given recent debate about marginally resolved gravitational-wave detection claims, the question of significance has practical consequences. We propose a Bayesian framework for calculating the odds that a signal is of astrophysical origin versus instrumental noise without bootstrap noise estimation. We show how the astrophysical odds can safely accommodate glitches. We argue that it is statistically optimal. We demonstrate the method with simulated noise and provide examples to build intuition about this new approach to significance.

\end{abstract}

\maketitle

\section{Introduction}
\label{sec:intro}
Recent breakthroughs in gravitational-wave astronomy have been facilitated by Bayesian inference \citep{veitch2015}.
Inference allows robust and unbiased determination of the source properties, model comparisons, population-level inferences for a set of observations, and optimal searches for the gravitational wave background (for examples, see, \citep{gw150914_properties, 2018PhRvL.121p1101A, soumi_2018, 2018arXiv181112940T, smith2018, known_pulsars, pitkin2017, powell2017}). However, the problem of detection for transient compact binary coalescence signals has typically been cast in frequentist terms.

Detection is the problem of identifying and quantifying the significance of astrophysical signals in noisy data.
To first order, the noise in gravitational-wave observatories like LIGO/Virgo~\citep{LIGO, virgo} can be described as a colored Gaussian process.
However, gravitational-wave observatories are subject to frequent transient artifacts known as \emph{glitches}; see, for example, \citep{2008CQGra..25r4004B, cabero2019, powell2018}.
Glitches can arise from any number of reasons, for example, photodiode saturation, environmental influence, and scattered light~\cite{O1_detchar_paper}.
In most cases, however, the cause of a glitch is unknown.
Whatever their source, glitches are not believed to be causally connected between sites.
That is, glitches in different observatories are independent and any coincidences are due entirely to chance.

For the gravitational-wave transients detected thus far, a $p$-value approach (also known as null-hypothesis significance testing) is often applied to assign significance for detection claims (see, e.g., \citep{gw150914_detection, GW150914_CBC_COMPANION, GW150914_BURST_COMPANION, 2016CQGra..33m4001A}).
A detection statistic for the signal is compared against a background distribution.
Without the ability to shield the observatory from astrophysical signals, the background distribution is generated instead using bootstrap methods such as time-slides (see, e.g. \citep{S3S4_inspiral, S4Burst, PyCBC, SPIIR, QiChuThesis, 2016PhRvD..94d4050L, 2017PhRvD..95j4046L, cwb}).
The term ``bootstrap'' refers to the practice of using an empirical distribution (in this case, a detection statistic) to estimate the properties of an estimator (in this case, the $p$-value); for more information, see, e.g.,~\cite{Efron}.

Time-slides are a classic example.
Data from independent observatories is time-shifted by an amount greater than the coherence time of the signal.
Each time shift, provides an independent realization of bootstrap generated noise.
By calculating the detection statistic for each time shifted data set, a background distribution can be generated. Astrophysical signals are detected when their detection statistic is suitably large compared to this background distribution. Time slides are a highly successful and convenient way to detect signals in the presence of transient noise. However, they have limitations.

With O3 sensitivity, the rate of binary black hole detections is expected to average 1 per week \citep{2018arXiv181112940T, 2018LRR....21....3A}.
If several days of data around the event are used to perform background estimation, there is non-negligible probability of including another event close to the detection threshold.
Using data containing astrophysical signals violates the basic assumptions of the time-slide method and will eventually (with increasing sensitivity) result in false dismissal of astrophysical signals due to contamination from signals in the background estimation procedure \citep{bigdog}.
In addition, it was shown by \citet{was_2010} that the time-slide method suffers a saturation problem: increasing the number of time-slides eventually fails to produce new realisations of the background data due to the limited variety of glitches in the original dataset.
This means the $p$-value estimated from time-slides can be incorrect and will not be improved by more time-slides~\citep{was_2010}.

An alternative bootstrap method to time-slides is to model the distribution of noise as a Poisson process, which combines random draws from single-detector background distributions~\citep{GSTLAL_D,GSTLAL_E}.
The background distribution is modeled using the empirical distribution of the single-detector detection statistic.
Employing bootstrap techniques, the method is subject to the same limitations as time-slides: saturation (from the limited noise realization available for the empirical distribution) and contamination from signals.
See~\citep{GSTLAL_A, GSTLAL_B, GSTLAL_C} for recent updates and applications of this method. 

Large-scale mock data challenges comparing pipelines \citep{capano2017} have validated the $p$-value approach to significance for rare events (unambiguous detections) using various bootstrap methods.
However, the authors of~\citep{capano2017} point out that bootstrap methods are subject to limitations when applied to marginal events\footnote{
As an aside, any $p$-value approach faces a fundamental limitation: the $p$-value cannot be used as a ``score'' to determine a candidate's significance.
As an example, GW150914 \citep{gw150914_detection} has a $p$-value, which is about $10^{5}$ times smaller than that of GW151012 \citep{gwtc1}.
Both of these pass the threshold for detection and therefore are regarded as detections.
However, to compare their $p$-values as a measure of how ``astrophysical'' they are is to fall for the transposed conditional fallacy.}.

These limitations motivate a fully Bayesian approach which eschews the bootstrap estimation of the background~a conclusion supported by \citep{capano2017}.
We propose an approach, which unifies, in a single framework, the problems of significance and parameter estimation.
We argue that the significance of a candidate event does not depend on the detection pipeline(s) used to identify it.

We anticipate that this Bayesian approach has broader applications for population modelling, robust multi-messenger detection \citep{2018ApJ...860....6A}, and detector characterization. 

There has been some work already toward this end.
The first LIGO/Virgo gravitational-wave transient catalog~\cite{gwtc1} includes a table of $p_\text{astro}$, a Bayesian odds comparing the astrophysical/terrestrial hypotheses~\citep{FGMC, Kapadia2019, digging}.
We regard this step as an important development.
The method for calculating $p_\text{astro}$ is fully Bayesian.
However, it relies on the bootstrap noise models of the search pipelines used to identify the candidate events.
Thus, the limitations of bootstrap methods discussed above affect the determination of $p_\text{astro}$.
Moreover, using the bootstrap noise models of different search pipelines leads to another undesirable consequences which is highlighted in Table~IV of~\cite{gwtc1}: different pipelines produce different values of $p_\text{astro}$.
Consider, for example, GW170729: the $p_\text{astro}$ values range from 52-98\%.
Recent claims from~\cite{ias} have yielded new candidates---using an entirely different pipeline---which were not deemed significant by LIGO/Virgo, at least in the first gravitational-wave transient catalog~\cite{gwtc1}.
This begs the question: what is the {\em actual} probability that GW170729 is of astrophysical origin?
We argue that the question is best answered in a pipeline-independent way, using the same infrastructure used for parameter estimation.

The remainder of this paper is organized as follows. In Sec.~\ref{sec:towards}, we introduce the formalism for calculating the \textodds, quantifying the probability that a data segment contains a signal of astrophysical origin versus noise. In Sec.~\ref{sec:simulation_knowing_the_odds}, we carry out a toy model demonstration to build intuition for the odds before providing a full-scale simulation study of binary black hole signals in modelled interferometer data in Sec.~\ref{sec:simulation_bbh}. We conclude in Sec.~\ref{sec:discussion} with a discussion and future outlook.

\section{Toward Bayesian detection}
\label{sec:towards}
In this section, we introduce a Bayesian odds for determining the significance of a gravitational-wave candidate without time slides.
The details are somewhat technical and it is necessary to introduce a bit of notation.
Rather than build the odds from constituent parts, we opt to begin with the result and then explain the components one at a time.

This section is organized as follows.
In~\ref{sec:formalism}, we provide a general expression for the \textodds and define the component parts.
In~\ref{sec:mixture}, we introduce a mixture model describing the data as a combination of signal and noise.
In~\ref{sec:noise}, we describe the noise model, which includes a model of glitches.
In~\ref{sec:signal}, we describe the signal model for gravitational waves from compact binaries.
In~\ref{sec:result}, we provide a practical expression for the \textodds based on our signal and noise models.
Finally, in~\ref{sec:literature}, we place the method in the context of previous literature.

\subsection{Formalism}\label{sec:formalism}
We presuppose that the data are divided into segments, each of which may contain a single binary signal.
The following boxed equation is the \emph{\textodds}, which answers the question ``what are the odds that the $i$th data segment contains a signal $\Hs_i$, given all the data $\alldata$?''
\begin{align}
\boxed{
\odds_{\Hs_i / \Hn_i}(\alldata) =
\frac{\int d\Lambda\, {\cal L}(\data_i| \Hs_i, \Lambda) \pi(\Lambda | \alldata_{i\neq k}) \pi(\Hs_i|\alldata_{i\neq k},\Lambda)}
{\int d\Lambda\, {\cal L}(\data_i| \Hn_i, \Lambda) \pi(\Lambda | \alldata_{i\neq k})\pi(\Hn_i|\alldata_{i\neq k},\Lambda)} .
\label{eqn:odds}
}
\end{align}
Let us go through component pieces in turn:
\begin{itemize}
    \setlength{\itemsep}{1pt}
    \item $\odds_{\Hs_i / \Hn_i}(\alldata)$. The detection statistic is a \emph{hyperparameterized astrophysical odds}, or for short, \emph{\textodds}, a single number, which we denote $\odds$. See Appendix~\ref{sec:hyper-odds} for a derivation of the hyper-odds in general. Like all odds, it compares two hypotheses, in this case: $\Hs$, the {\em signal hypothesis}, and $\Hn$, the {\em noise hypothesis}. According to the signal hypothesis, the segment contains an astrophysical signal in amongst noise. According to the noise hypothesis, there is no signal present, just noise. Noise, here, refers to both Gaussian noise and/or a glitch. The \textodds depends on the {\em full dataset}~$\alldata$.
    \item $\Lambda$. The numerator and denominator include integrals over $\Lambda$, which is a set of {\em hyperparameters} modelling the  distribution of {\em signal and noise parameters} $\theta_i$. The models are described using {\em conditional priors} $\pi(\theta_i|\Lambda)$.
    \item $d_i$ and $\alldata_{i\neq k}$. We divide $\alldata$ into $d_i$, the $i$th data segment, and every other segment of data, $\alldata_{i\neq k}$, which we refer to as the \emph{contextual data}. The contextual data provides context with which to understand the significance of $d_i$.
    \item ${\cal L}(\data_i| \Hs_i, \Lambda)$. The next term in the numerator is the likelihood of the data given the signal hypothesis, and given the hyperparameters. For short, we call it {\em the signal evidence}. The corresponding term in the denominator is ${\cal L}(\data_i| \Hn_i, \Lambda)$, the likelihood of the data given the noise hypothesis, and given the hyperparameters. For short, we call it {\em the noise evidence}.
    \item $\pi(\Lambda | \alldata_{i\neq k})$. The next term in the numerator is the posterior for the hyperparameters given the contextual data. We use this $i\neq k$ posterior as the prior for event $i$. For short, we call this {\em the hyper-prior}.
    \item $\pi(\Hs_i|\alldata_{i\neq k},\Lambda)$. The final term in the numerator is the prior for the signal hypothesis given the contextual data and given the hyperparameter $\Lambda$. For short,we call this {\em the signal prior}. The corresponding term in the denominator is the prior for the noise hypothesis, $\pi(\Hn_i|\alldata_{i\neq k},\Lambda)$. For short, we call this {\em the noise prior}.
\end{itemize}

Qualitatively, Eq.~\eqref{eqn:odds} is straightforward to understand now that we have introduced the necessary notation.
Our signal and our noise are described by parameters $\theta_i$.
The prior distribution for $\theta_i$ has an uncertain shape, which we model using hyperparameters $\Lambda$, so that the signal and noise parameters are conditional on the hyperparameters: $\pi(\theta_i|\Lambda)$.
In other words, we employ a hierarchical model. The odds, defined in Eq.~\eqref{eqn:odds}, are an example of a hyper-odds (see Appendix~\ref{sec:hyper-odds}) which marginalizes over uncertainty in the hyperparameters using contextual data $\alldata_{i\neq k}$. The overall result is that Eq.~\eqref{eqn:odds} compares the probability for a signal and noise model evaluated using a hierarchical model conditional on the contextual data.
In the following sub-sections, we will describe how each term in Eq.~\eqref{eqn:odds} is calculated in detail, resulting in the version used in practise, Eq.~\eqref{eqn:odds2}.



\subsection{Mixture model}\label{sec:mixture}
In Sec.~\ref{sec:formalism} we presented the \textodds in a general form.
It is now necessary to specialize further.
First, we assume that the hyperparameters consist of three components:
\begin{align}
    \Lambda \rightarrow & \{\Lambda_\Hs, \Lambda_{\Hnots}, \xi\} \\
    = & \{\Lambda_{\Hnots}, \xi\}\,,
\end{align}
where $\xi \equiv \pi(\Hs| \xi)$ is the mixing fraction and 
the signal hyperparameters $\Lambda_\Hs$ determine the shape of the signal priors.
For this analysis, we assume there are no signal-hyper parameters and so this term can be neglected:
\begin{align}
    {\cal L}(\data_i| \Hs_i, \Lambda) \rightarrow & {\cal L}(\data_i| \Hs_i, \Lambda_\Hs) \\
    = & {\cal L}(\data_i| \Hs_i).
\end{align}
However, one could straightforwardly extend the analysis to incorporate signal hyperparameters, for example, from~\cite{O2R&P}.

The noise hyperparameters $\Lambda_{\Hnots}$ determine the shape of the noise prior:
\begin{align}
    {\cal L}(\data_i| \Hnots_i, \Lambda) \rightarrow & {\cal L}(\data_i| \Hnots_i, \Lambda_{\Hnots} ) .
\end{align}
This is where the action happens: the problem of determining the significance of a candidate event is recast as a problem of determining a suitable hyperparameterization for the noise distribution.

Constructing a mixture model for the exhaustive hypothesis $\Hs \lor \Hn$, we obtain a {\em general hyperlikelihood}, conditional on the hyperparameters:
\begin{align}
    {\cal L}(d | \Lambda_\Hs, \Lambda_{\Hnots}, \xi, \Hs \lor \Hn) =
    \xi {\cal L}(d | \Hs) + (1-\xi){\cal L}(d | \Lambda_{\Hnots}, \Hn)\,.
    \label{eqn:mixture_likelihood}
\end{align}
The general likelihood is used to calculate $\pi(\Lambda | \alldata_{i\neq k})$.
In Section~\ref{sec:noise} and Section~\ref{sec:signal}, we give details on how the noise and signal evidences are calculated in practise.

\subsection{Noise model}\label{sec:noise}
We adopt a noise model consisting of multiple sub-hypotheses and for simplicity restrict the discussion to just two observatories $a$ and $b$.
The entire noise hypothesis, which is the union of four sub-hypotheses, is denoted $\Hn$.
Each sub-hypothesis accounts for the different kinds of noise.
For example, $\Hnots_0\Hg^{a}\Hnotg^{b}$ is the sub-hypothesis that: no signal is present ($\Hnots_0$); there is a glitch in observatory $a$ ($\Hg^a$); there is no glitch present in observatory $b$ ($\Hnotg^b$).
The complete noise hypothesis is
\begin{align}
\Hn \equiv \Hnots_0\Hg^{a}\Hg^{b} \lor
           \Hnots_0\Hg^{a}\Hnotg^{b} \lor
           \Hnots_0\Hnotg^{a}\Hg^{b} \lor
           \Hnots_0{\Hnotg^{a}}\Hnotg^{b}\,.
\label{eqn:Hn}
\end{align}
The noise evidence from Eq.~\eqref{eqn:mixture_likelihood} can therefore be calculated using a mixture model
\begin{align}
\begin{split}
    {\cal L}(d| \Lambda_{\Hnots_0}, \Hn) = &
    \xi_g^a \xi_g^b {\cal L}(d| \Lambda_{\Hnots}, \Hnots_0\Hg^a\Hg^b) \\
    & + \xi^g_a (1 - \xi^g_b){\cal L}(d| \Lambda_{\Hnots}, \Hnots_0\Hg^a\Hnotg^b) \\
    & +(1-\xi_g^a) \xi_g^b  {\cal L}(d| \Lambda_{\Hnots}, \Hnots_0\Hnotg^a\Hg^b) \\
    & + (1-\xi_g^a)(1-\xi_g^b) {\cal L}(d| \Lambda_{\Hnots}, \Hnots_0\Hnotg^a\Hnotg^b)\,,
\end{split}
\label{eqn:noise_likelihood}
\end{align}
where $\xi_g^x\equiv \pi(\Hg^x| \Lambda_{\Hnots})$ is the glitch mixing fraction for observatory $x$. The individual likelihoods in this expression are the evidence for glitches (or a lack thereof) in each observatory. Assuming the noise (including glitches) are independent between the two observatories, we can simplify the likelihoods, e.g.,
\begin{align}
    {\cal L}(d| \Lambda_{\Hnots}, \Hnots_0 \Hg^a\Hg^b) = {\cal L}(d| \Lambda_{\Hnots}, \Hnots_0 \Hg^a){\cal L}(d| \Lambda_{\Hnots}, \Hnots_0 \Hg^b)\,.
    \label{eqn:glitch_independence}
\end{align}

In the absence of a signal or glitch hypothesis, e.g., ${\cal L}(\data| \Hnots \Hnotg^x)$, the evidence is the usual Gaussian-noise evidence \citep{thrane2019}. On the other hand,
given a glitch model, the evidence for a glitch in the $x$-observatory is calculated from marginalizing over the glitch model-parameters
\begin{align}
    {\cal L}(\data| \Lambda_{\Hnots}, \Hnots \Hg^x) = \int d\theta\,
    {\cal L}(\data| \Lambda_{\Hnots}, \theta, \Hg^x) \pi(\theta| \Lambda_{\Hnots}, \Hg^x)\,.
    \label{eqn:glitch_evidence}
\end{align}
In practise, this integration is performed numerically using nested sampling methods \citep{skilling2004}. In order to allow rapid evaluation of ${\cal L}(\data| \Lambda_{\Hnots}, \Hnots \Hg^x)$ while varying $\Lambda_{\Hnots}$, we use the so-called recycling method; see Appendix~\ref{sec:recycling} and Refs.~\citep{thrane2019, talbot2018, pitkin2018}.

However, it is difficult to build a glitch model to evaluate Eq.~\eqref{eqn:glitch_evidence} from first principles since most glitches are poorly understood.
In this work, we instead apply the conservative model first proposed by \citet{veitch2010} in which glitches are modeled as compact binary signals with uncorrelated model parameters in each observatory. 

This glitch model is founded on the principle of modelling the worst-case scenario: glitches are indistinguishable from signals except for the absence of coherence (in the model parameters $\theta)$ between observatories. Therefore for a signal to be preferred over this glitch model, it must not only match modelled waveforms, but must also look like the same binary system in multiple observatories with arrival times consistent with an astrophysical origin.
We will apply this idea, further developed in \citet{isi2018},
which extends the noise hypothesis to include Gaussian noise.

This glitch model is conservative in that we could better distinguish glitches and hence boost the significance of astrophysical signals by including more physically motivated models of glitches (see, e.g., Ref.~\citep{cornish2015}). 
However, in the absence of a trustworthy physical glitch model, our conservative approach ensures that we do not generate false positives due to a flaw in our glitch model.
That said, this formalism can be extended to accommodate more sophisticated models. We return to this below when discussing possibilities for future work.

\subsection{Signal model}\label{sec:signal}
We adopt a signal model consisting of multiple sub-hypotheses and for simplicity restrict the discussion to just two observatories $a$ and $b$.
The entire signal hypothesis, which is the union of four sub-hypotheses, is denoted $\Hs$.
Each sub-hypothesis accounts for the different kinds of noise that the signal can be embedded within.
For example, $\Hs_0\Hg^{a}\Hnotg^{b}$ is the sub-hypothesis that a signal is present ($\Hs_0$) with a glitch in observatory $a$ ($\Hg^a$) with no glitch present in observatory $b$ ($\Hnotg^b$). 
With these definitions, the complete signal hypothesis is 
\begin{align}
\Hs \equiv \Hs_0\Hg^{a}\Hg^{b}
           \lor \Hs_0\Hg^{a}\Hnotg^{b}
           \lor \Hs_0 \Hnotg^{a}\Hg^{b}
           \lor \Hs_0 {\Hnotg^{a}}\Hnotg^{b}\,.
\label{eqn:Hs}
\end{align}
Qualitatively, this expression captures the idea that a signal in the detector $\Hs_0$ may coincide with or without a glitch in either detector.
Note that, while we refer to $\Hs$ as ``the signal hypothesis'' for brevity, it might be more aptly named ``the set of all hypotheses that include a signal.''

Signals from compact binaries are typically characterized by 15 parameters.
In the previous subsection, we introduced a noise model where glitches modeled as incoherent binary signals, which introduces 15 parameters per observatory. 
Thus, this formulation will require marginalizing over 30 parameters (for the $\Hs\Hg^a\Hnotg^b$, $\Hs\Hnotg^a,\Hg^b$ sub-hypotheses) and marginalizing over a 45-dimensional parameter space for the $\Hs\Hnotg^a\Hnotg^b$ sub-hypothesis. With current nested-sampling
methods, these integrals remain challenging and robust methods to calculate them are an unsolved problem which has a number of applications beyond the current work.

However, if the rate of astrophysical signals is sufficiently small, then we can make the following approximation as in ~\citet{smith2018}:
\begin{align}
\Hs \approx \Hs_0 {\Hnotg^{a}}\Hnotg^{b}\,.
\label{eqn:Hs_truncated}
\end{align}
As such, the signal evidence in Eq.~\eqref{eqn:mixture_likelihood} is calculated by simply marginalizing over binary parameters
\begin{align}
    {\cal L}(\data| \Hs) = {\cal L}(\data| \Hs_0 {\Hnotg^{a}}\Hnotg^{b}) = \int d\theta
    {\cal L}(\data| \theta, \Hs_0 ) \pi(\theta| \Hs_0)\,,
    \label{eqn:signal_evidence}
\end{align}
where we drop the conditional probability on the no-glitch hypotheses where they have no bearing in the signal likelihood, ${\cal L}(\data_i| \theta, \Hs_0)$. The signal likelihood is constructed from a Gaussian noise likelihood (with the power spectral density estimated from the data) and standard stochastic sampling procedures are used to perform the marginalization (see, e.g., \citep{thrane2019} for detailed discussion).

\subsection{The \textodds: a practical implementation}\label{sec:result}
The \textodds have been given in the generic form of Eq.~\eqref{eqn:odds}. Having defined the noise and signal models (see Sec.~\ref{sec:noise}, ~\ref{sec:signal}), we obtain an expression for our specific signal and noise models:
\begin{align}
\boxed{
\odds_{\Hs_i / \Hn_i}(\alldata) =
\frac{{\cal L}(\data_i| \Hs_i) \int d\xi \, \xi \pi(\xi | \alldata_{i\neq k})}
{\iint d\xi d\Lambda_{\Hn} {\cal L}(\data_i| \Hn_i, \Lambda_{\Hn}) (1-\xi) \pi(\xi, \Lambda_{\Hn} | \alldata_{i\neq k})} .
}
\label{eqn:odds2}
\end{align}
The odds of Eq.~\eqref{eqn:odds2} are the Bayesian answer to the question: does data segment $\data_i$ contain an astrophysical signal or is it noise? Where the noise hypothesis includes a conservative glitch model and the question is asked, not in the isolated case of a single data segment, but in the context of some wider set of data $\alldata$.

That the odds are a fully Bayesian answer is important.
There is no need to treat this odds as a frequentist detection statistic. There is no need to perform time-slides to generate a background and then calculate a false alarm rate, a technique which runs into problems with saturation. The odds are a statement about the probability that the data is a signal rather than noise; for example, an odds of 9:1 will be a signal nine times out of 10. We discuss this point further in Sec.~\ref{sec:simulation_knowing_the_odds}.

A point we have yet to discuss, and will be the core issue for practical applications of this method, is that the odds will be sensitive to the validity of the noise hyper-model. However, this is not a drawback to the method, but a feature.
This method makes explicit the underlying noise model which is being applied. This will require careful checking to ensure the noise hyper-model is appropriate using posterior predictive checks. 

\subsection{The astrophysical odds: relation to previous results}\label{sec:literature}
The hyperparameterisation step in the \textodds is unique to this work. However, our work builds on ideas in the literature. Here, we show how these are related.

In \citet{veitch2010}, the coherence test was introduced, which is a Bayes factor between a coherent signal $\Hs$ and incoherent glitches $\Hg^{a}\Hg^B$. In the notation of this work, this is
\begin{equation}
    B_\textrm{coh,inc}(\data_i) \equiv \frac{{\cal L}(\Hs | \data_i)}{{\cal L}(\Hg^a\Hg^b | \data_i)} = 
    \frac{{\cal L}(\data_i | \Hs)}{{\cal L}(\data_i| \Hg^a){\cal L}(\data_i| \Hg^b)}\,,
    \label{eqn:BCI}
\end{equation}
with $\Hs$ as defined in Eq.~\eqref{eqn:Hs_truncated}.

Later, \citet{isi2018} introduced the BCR-statistic,
\begin{equation}
    \textrm{BCR} \equiv \frac{{\cal L}(\Hs | \data_i)}{{\cal L}( \Hn | \data_i)} {=} \frac{\alpha {\cal L}(\data_i| \Hs)}
    {\displaystyle \prod_{\ell=a,b}\left[{\cal L}(\data_i| \Hg^\ell)\beta_\ell + (1-\beta_\ell){\cal L}(\data_i | \Hnotg^\ell)\right]}\,,
    \label{eqn:BCR}
\end{equation}
which is an odds comparing the signal hypothesis $\Hs$, as defined in Eq.~\eqref{eqn:Hs_truncated}, with incoherent glitches or Gaussian noise. Comparing with the work herein, $\alpha = \xi/(1-\xi)$ and $\beta_\ell = \xi_g^\ell$. In the case where no contextual data is used, the \textodds, Eq.~\eqref{eqn:odds2}, are a generalization of the BCR odds. \citet{isi2018} demonstrate how to the rate of signals and glitches ($\alpha$ and $\beta$) can instead be tuned using time-slid data and injections to maximally separate signals from noise. This is analogous to how $\xi$ and $\xi_g$ are estimated in the \textodds, but differs in that the work presented herein uses contextual non-time-slid data. The \textodds also allows the glitch model-parameter hyperpriors to be inferred from the contextual data.


\section{Toy model demonstration}
\label{sec:simulation_knowing_the_odds}

In this section we demonstrate the Bayesian odds with a toy-model problem.
The point of this exercise is to show that: (1) given well-informed priors, the Bayesian odds has a clear and reliable interpretation; (2) if the prior is misinformed (i.e. does not represent our true beliefs), this interpretation becomes unreliable. Later, in Sec.~\ref{sec:simulation_bbh}, we show that if the exact prior is unknown, it can be inferred using hierarchical modelling, reestablishing the reliability of the Bayesian odds.

In our toy model, each measurement consists of a single number $x_i$.
The data are generated either by a noise model, consisting of
a standard normal distribution (i.e., zero mean and unit variance), or a
signal mode, consisting of a normal distribution with some non-zero mean and
unit variance. 
To summarize, our signal and noise hypotheses are:
\begin{itemize}
\item{$\Hs$:} $x_i\sim \mathrm{Normal}(\mu, \sigma{=}1)$ with $\mu \sim \mathrm{Uniform}(0, 1)$
\item{$\Hn$:} $x_i\sim \mathrm{Normal}(\mu{=}0, \sigma{=}1)$
\end{itemize}
We fix the prior-odds to unity, i.e., $\pi(\Hs)=\pi(\Hn)=1/2$.

We simulate a data set consisting of 10 data points and randomly assign each one to the ${\cal S}$ and ${\cal N}$ categories. Next, we draw random values of $x_i$ based on the category of each event.
Having generated a set of data $\{x_i\}$, we calculate the evidences ${\cal L}(\{x_i\}|\Hn)$ and ${\cal L}(\{x_i\}|\Hs)$. The first of these can be calculated
directly, the second is estimated using a nested-sampling algorithm,
marginalising over $\mu$.
Initially, we apply the prior $\pi(\mu|
\Hs)=\mathrm{Uniform}(0, 1)$, i.e., the correct population prior used to generate the data.
(Below we repeat the calculation using an intentionally misinformed prior for illustrative purposes.)
Once these estimates of the evidence are calculated, we calculate the odds $\odds$ for the data set.

\begin{figure}[tb]
\centering
\includegraphics[width=\columnwidth]{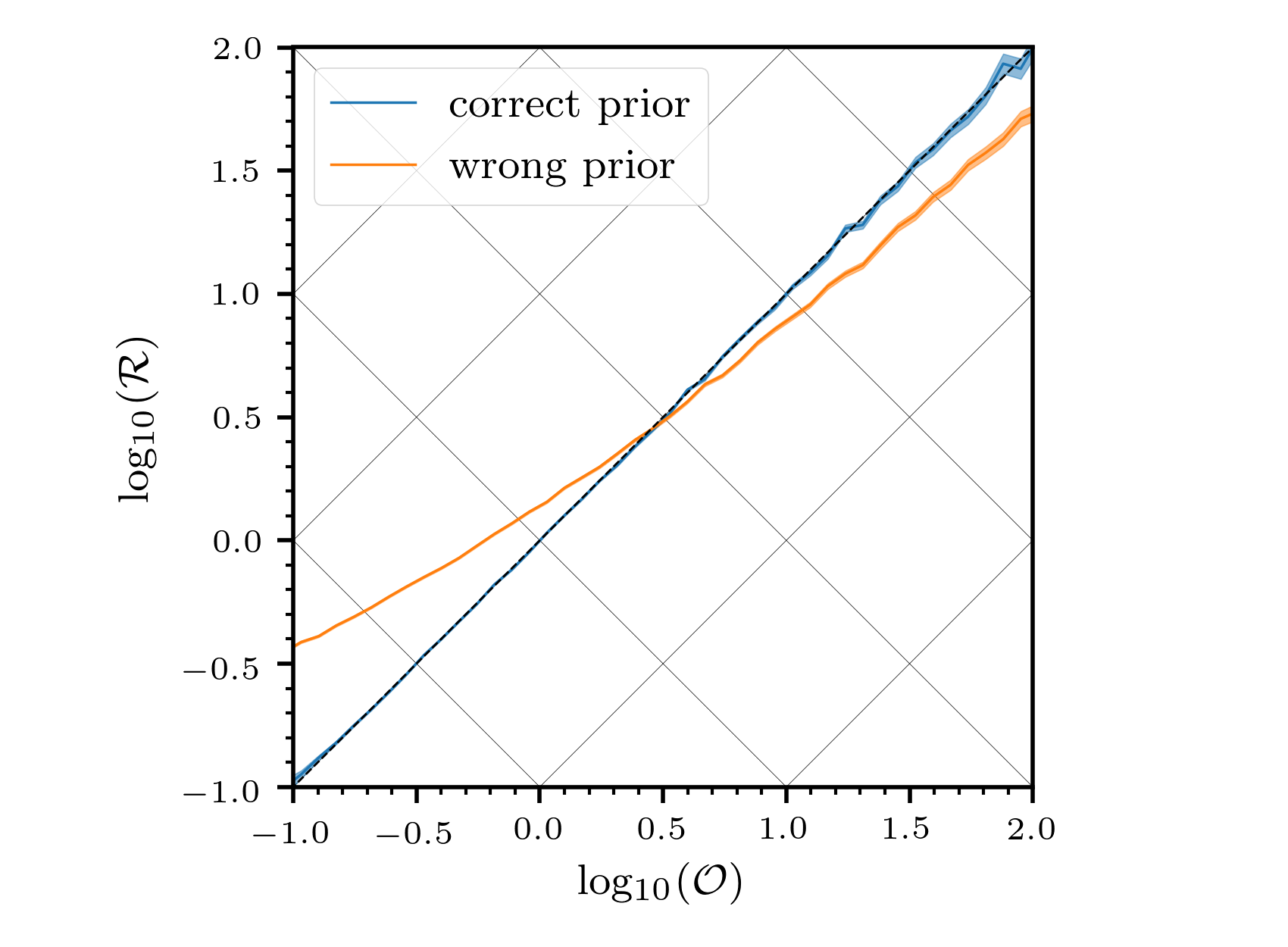}
\caption{Performance plot of the odds using the correct and incorrect
prior (see text). Shaded regions indicate the approximate $1{-}\sigma$ uncertainty
due to the finite number of realisations used. An optimally performing odds
lies on the diagonal line.}
\label{fig_toy_model}
\end{figure}

We demonstrate that if the odds are
formulated correctly, they can be trusted at face value. 
That is, events with odds of ${\cal O}=9$ will turn out to be signal nine times out of ten.
We illustrate this with a plot. 
Let $\delta M_{\Hs}(\odds)$ be the number of data-sets simulated as signals 
with an odds on the interval $(\odds, \odds + \delta\odds)$. Then,
if $\delta M_{\Hn}(\odds)$ is the number of simulated data-sets simulated as noise
on the same interval we can define
\begin{align}
\mathcal{R}(\odds) \equiv \frac{\delta M_{\Hs}(\odds)}{\delta M_{\Hn}(\odds)}\,.
\end{align}
If the odds are what they say they are, then this ratio should exactly
equal the odds, i.e., $\mathcal{R}=\odds$.
In Fig.~\ref{fig_toy_model}, we simulate 1000 data sets and plot $\log_{10}(\mathcal{R})$ against
$\log_{10}(\odds)$ using the \emph{correct prior}: i.e., the prior distribution
from which the model parameters where drawn (blue). The odds perform
as expected: the diagonal, corresponding to $\mathcal{R}=\odds$, is consistent with the diagonal ${\cal R}={\cal O}$ line.

For comparison, we can purposefully repeat the calculation using a \emph{misinformed prior} to understand
what effect this has on the plot. We repeat the
steps above, but when calculating the signal evidence, we do not use the
correct prior distribution on $\mu$, but instead use an intentionally misinformed prior $P(\mu| \Hs')
\propto \mu^{2}$ with a minimum of 0 and a maximum of 1, i.e. a power-law
distribution with the same support as the correct prior, but a different spectral
index. In effect, we are comparing a new hypothesis $\Hs'$ with the noise instead of $\Hs$.
The result is shown in Fig.~\ref{fig_toy_model} (orange). Unlike the case of the
correct prior, the odds now do not behave as expected: at times it is too
liberal and and other times too conservative.

\section{Demonstration with binary black hole mergers}
\label{sec:simulation_bbh}
We now simulate the problem of binary coalescence significance estimation with data from two observatories (labelled $H$ and $L$) containing binary black hole signals and glitches (as described in Sec.~\ref{sec:noise} and Sec.~\ref{sec:signal}). From this simulated data, we first infer the population hyperparameters, verifying that the values used in generating the simulation are properly recovered. Then, we demonstrate calculation of the \textodds, showing the behaviour as a function of the amount of contextual data. We use the \texttt{Bilby} \citep{bilby2019} Bayesian inference package to generate and analyse the simulated data and the \texttt{dynesty} \citep{dynesty} nested-sampling package for parameter estimation and evidence evaluation.

We define the simulated population parameters in Table~\ref{tab:sim_pop_params}. The signal and glitch rates ($\xi, \xi_g^H$, and $\xi_g^L$) refer to the probability of a \unit[4]{s} data segment containing a signal or glitch. Their values are selected so that in a sample of a few hundred data segments, the expected number of segments containing a signal and glitch is less than one, but the absolute number of signals and glitches is sufficiently large enough for population-level inference. Meanwhile, Table~\ref{tab:sim_pop_params} also outlines the population distribution of luminosity distances and chirp mass for signals and glitches; these have the same support, but obey different scaling-laws. The population hyperparameters are chosen so that glitches are more frequent and tend to be louder and shorter in duration than signals. 

\begin{table}[tb]
    \centering
    \caption{Binary black hole coalescence simulation population parameters. The signal rate, $\xi$,  and glitch rate for the $H$ and $L$ detectors, $\xi^H$ and $\xi^L$, refer to the probability of a \unit[4]{s} data segment containing a signal or glitch. The luminosity distance (chirp-mass) population distributions for signals, $d_L^s$ ($\mathcal{M}_s$) and glitches $d_L^g$ ($\mathcal{M}_g$) have equal support, but obey differing scaling relations; the same glitch population is applied to both the $H$ and $L$ observatories.}
    \label{tab:sim_pop_params}
    \begin{tabular}{rl}
         Parameter & \;\;\; Distribution \\ \hhline {==}
         $\xi$ & = 0.001 \\
         $\xi_{g}^H$ &= $0.6$\\
         $\xi_g^L$ & = $0.4$ \\
         $d_L^s$ & $\sim \mathrm{PowerLaw}(\alpha{=}2, \unit[1]{Gpc}, \unit[2]{Gpc})$ \\
         $d_L^g$ & $\sim \mathrm{PowerLaw}(\alpha{=}0, \unit[1]{Gpc}, \unit[2]{Gpc})$ \\
         $\mathcal{M}_s$ & $\sim \mathrm{PowerLaw}(\alpha{=}0, \unit[25]{M_{\odot}}, \unit[100]{M_{\odot}}$ \\
         $\mathcal{M}_g$ & $\sim \mathrm{PowerLaw}(\alpha{=}2, \unit[25]{M_{\odot}}, \unit[100]{M_{\odot}}$
    \end{tabular}
\end{table}

Each segment of the contextual data is generated by first drawing parameters from the population level rate parameters (i.e., Table~\ref{tab:sim_pop_params}) to determine what the segment should contain.
If the segment is to contain a signal, a set of signal parameters are drawn from a standard set of priors, but with the luminosity distance and chirp mass as given in Table~\ref{tab:sim_pop_params}.
If the segment is to contain a glitch, a set of glitch parameters (for each observatory) are drawn from a standard set of priors identical to the standard signal priors, but with the luminosity distance and chirp mass as given in Table~\ref{tab:sim_pop_params}. Afterwards, the simulated signals and glitches are added to Gaussian noise. The simulation will add both signals and glitches to the data simultaneously, however we have chosen the rate parameters to ensure the probability of this occurring is small.

\begin{figure}
    \centering
    \includegraphics[width=\columnwidth]{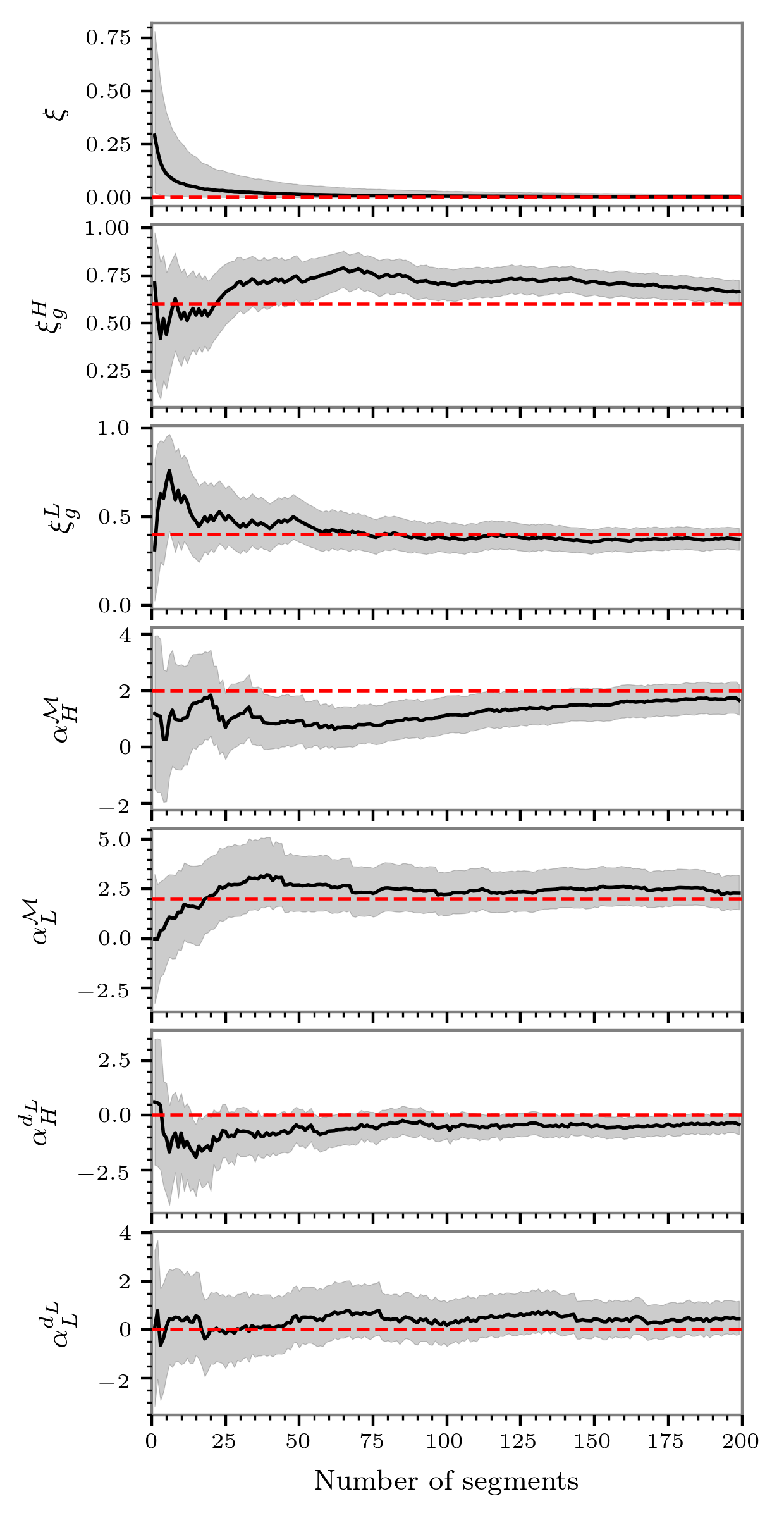}
    \caption{Marginalized posterior median (solid line) and 90\% credible interval (shaded region) for the hyperparameter of the binary black hole simulation. True values are shown as horizontal dashed lines.}
    \label{fig_conv}
\end{figure}

Having generated the contextual data segments, we proceed to recover the hyperparameters; this is done by applying Eq.~\eqref{eqn:mixture_likelihood}, a hyperparameter model for each observatory is used for both the luminosity distance and chirp mass. Each model has the same support as the true priors (see Table~\ref{tab:sim_pop_params}), but an unknown spectral index $\alpha$ with subscript either $H$ or $L$, labelling the observatory and superscript labelling if it is for the luminosity distance $d_L$ or chirp mass $\mathcal{M}$. 
We repeat the hyperparameter inference for a variable number of segments randomly drawn from the prior. In Fig.~\ref{fig_conv}, we show the marginalized one-dimensional posterior distributions for the hyperparameters as a function of the number of segments. This demonstrates the expected results that, as the number of segments increases, we correspondingly see the posteriors converge to the true values.

As the amount of contextual data increases, Fig.~\ref{fig_conv} demonstrates that inferences about the hyperparameters also become increasingly well informed. How does this affect the \textodds? To study this, we simulate a glitch in Gaussian observatory noise and compute the \textodds using Eq.~\eqref{eqn:odds} while varying the set of contextual data (using the same data sets used to produce Fig.~\ref{fig_conv}). The glitch is the worst possible type of glitch: a quasi-coherent glitch. Such a glitch consists of a binary black hole signal injected into both observatories with near-identical signal parameters to within the uncertainties afforded by the background Gaussian observatory noise. For false alarms, this is precisely the sort of glitch which is best at deceiving traditional detection statistics. 

The quasi-coherent glitch has a network optimal SNR of $\sim50$ and log-Bayes factor comparing a signal versus Gaussian noise of $\sim1000$; when compared to Gaussian noise it is distinctly signal like. However, the simulated contextual data contains a multitude of non-Gaussian artifacts. For illustrative purposes, we calculate the significance several different ways. The coherence test from \citep{veitch2010} yields a Bayes factor $\log B_\mathrm{coh, inc}=8.9$ (see Eq.\eqref{eqn:BCI} for a definition).
This is unsurprising given we injected a quasi-coherent glitch, which is essentially indistinguishable from a coherent signal. The BCR-method \citep{isi2018}, see Eq.\eqref{eqn:BCR} for a definition, depends on the tuning parameters $\alpha$ and $\beta$. Using uniformed values of $\alpha=1$ and $\beta=0.5$, we have that $\log \mathrm{BCR}=8.9$ .
In this case the alternative ``Gaussian noise'' hypothesis provided by the BCR \citep{isi2018} does not further distinguish the event. That these methods fail to veto this quasi-coherent glitch is unsurprising and precisely the motivation for the \textodds, which incorporates knowledge about the background. We note that tuning the BCR parameters would improve the performance of this metric, but is beyond the scope of this paper.

\begin{figure}[tb]
    \centering
    \includegraphics[width=\columnwidth]{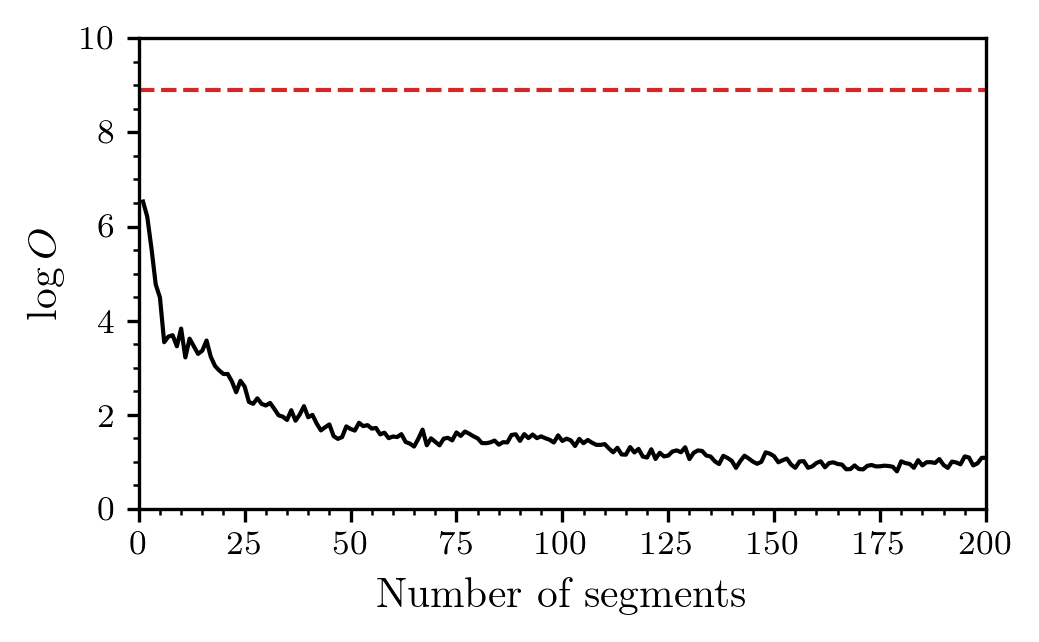}
    \caption{The \textodds, Eq.~\eqref{eqn:odds2} as a function of the number of contextual data segments used in the calculation of the background prior. A dashed horizontal line marks the value of the odds without any contextual data.}
    \label{fig:scurve}
\end{figure}

In Fig.~\ref{fig:scurve}, we plot the \textodds, for this same pathological event, as a function of the number of background segments used. For small numbers of segments, the astrophysical odds agrees with the $\log B_\mathrm{coh, inc}$ and $\log \mathrm{BCR}$ numbers, which makes sense since the background is not yet well-constrained. As the number of segments increases, however, the odds tend to decrease, eventually reaching a threshold of $\log\odds \approx 0$. In this instance, the \textodds are (appropriately) far more conservative than the other approaches. Given the nature of the background, the statistical significance of the event is marginal. It is worth noting that a similarly loud signal would also fail to pass the bar for detectability: fundamentally, the background in this region includes so many loud glitches that it decreases the operational sensitivity. 

That the odds tend to a threshold is a direct result of the limited amount of information available from the background as seen in Fig.~\ref{fig_conv}. Initially, each extra set of segments provides a substantial improvement to the constraint on the hyperpriors, but eventually once the background has been adequately sampled, extra data does not provide much new information about the nature of the background. The result is that one needs to make sure the number of background segments is sufficiently large before drawing a conclusion.
By studying this behaviour, gravitational-wave astronomers can ascertain how much data is required to determine the significance of a candidate before reaching the point of diminishing returns.

\section{Discussion and outlook}
\label{sec:discussion}
The tools of Bayesian inference have been widely adopted in astrophysics due to their utility in parameter estimation and model selection.
In this work, we introduce a complete formalism for calculating the \textodds (Eq.~\eqref{eqn:odds}) for transient gravitational wave signals. Gravitational wave observatories are plagued by transient non-Gaussian artifacts often referred to as glitches. These glitches need to be carefully considered when evaluating the significance of a putative signal.

Building on the work of \citet{veitch2010, isi2018, smith2018}, we present a complete framework for defining the likelihood of the data containing a signal, accounting for both Gaussian noise and glitches in the data. The conservative glitch model used herein relies on the ideas of pairs of incoherent compact binary signals as a conservative glitch model as proposed by \citet{veitch2010}. However, in the future it may prove fruitful to extend this glitch model to include, for example, sine-Gaussians \citep{cornish2015} in addition to compact binary signals.
If the data supports the hypothesis that glitches look more like sine-Gaussians than compact binaries, the \textodds will automatically adopt the less conservative assumption.

The key new ingredient developed in this work is the ability to marginalize over the glitch population-properties using contextual data. This is done using hyperparameters, which allow the glitch population to differ from the priors for astrophysical signals. For example, astrophysical signals are expected to follow a prior in which more signals are found at larger distance (in this paper, we model this as $P(d_L |I) \propto d_L^2$). But for glitches in the observatory (as described by fitting incoherent CBC signals), this is unlikely to be true.
Eq.~\eqref{eqn:odds} is the main result of this paper. It describes how the odds of an astrophysical signal can be calculated, marginalized over the contextual data which surrounds the event. This result is fully Bayesian: it does not require boot straps and hence does not suffer from issue such as saturation \citep{was_2010} or signal contamination. Moreover, the Bayesian approach (see also Refs.~\citep{FGMC, Kapadia2019, digging}) allows the significance to be easily integrated into downstream analyses, for example in population modelling.

One subtlety to this type of analysis is the characterisation of the background. We demonstrated in this work how injected background with known power-law distributions can be recovered. But, for real interferometer data, more nuanced models may be needed. In future work, we aim to apply this method to data from the LIGO and Virgo interferometers.
As part of this work, we aim to recompute the significance of previously published gravitational-wave detections and/or candidates~\cite{gwtc1,ogc1,ias}.
We plan to carry out diagnostic tests, for example, with posterior predictive checking, in order to build confidence in our noise model.

\section*{Acknowledgements}
The authors are grateful to Maximiliano Isi and members of the LIGO and Virgo compact binary coalescence group for review and useful comments during the preparation of this work.
GA, RS, and ET are supported by the Australian Research Council through grants CE170100004,  FT150100281, and DP180103155.  The authors gratefully acknowledge the support of the NSF, STFC, MPS, INFN, CNRS and the State
of Niedersachsen/Germany for provision of computational resources. This is document LIGO-P1900162.

\bibliography{bibliography}

\appendix

\section{Hyper-odds}
\label{sec:hyper-odds}
Typically, Bayesian model selection problems answer the question ``given some data $d$, what are the relative probabilities of model $A$ and model $B$?'' by calculating the odds, $\odds_{A/B}(d)\equiv P(A|d) / P(B|d)$. In this work, we answer the question
``given a set of data $\alldata$, what are the relative probabilities of model $A$ and model $B$ for the $i$th data segment?'' by calculating the generalised notion of a \emph{hyper-odds} which we define now. The important distinction here is that the data $\alldata_{k\neq i}$, which we refer to as \emph{contextual data}, is used to inform the odds about the typical characteristics of the $A$ and $B$ models.

Let $A_i$ be the hypothesis of model $A$ for the $i$th segment. Then, the model evidence for $A_i$ given all of the data can be calculated from
\begin{align}
{\cal L}(A_i | \alldata) & = \int d\Lambda\,
{\cal L}(A_i | \alldata, \Lambda)\pi(\Lambda | \alldata)\,.
\end{align}
Applying the rules of conditional probability and explicitly writing the likelihood conditional on the data for segment $i$ and for the contextual data, we find that
\begin{align}
{\cal L}(A_i | \alldata) = \frac{1}{{\cal L}(\alldata)}\int d\Lambda\,
{\cal L}(\data_i| A_i, \Lambda) \prod_{k\neq i}{\cal L}(\data_k| \Lambda) \pi(\Lambda)\,.
\label{eqn:hyper-evidence-A}
\end{align}

Typically, the normalising factor in this equation cannot easily be
calculated since an exhaustive set of models is seldom known. Instead, the common alternative is to instead calculate an odds
\begin{align}
\begin{split}
\odds_{A_i / B_i}(\alldata) & =
\frac{{\cal L}(A_i| \alldata)}
{{\cal L}(B_i| \alldata)} \,.
\end{split}
\end{align}
Applying Eq.~\eqref{eqn:hyper-evidence-A}, we therefore have that
\begin{align}
\odds_{A_i/B_i}(\alldata){=}
\frac{
\int d\Lambda
{\cal L}(\data_i| \Lambda, A_i)
{\cal L}(A_i|\Lambda)
\prod_{i\neq k}{\cal L}(\data_k| \Lambda)
\pi(\Lambda)
}{
\int d\Lambda
{\cal L}(\data_i| \Lambda, B_i)
{\cal L}(B_i|\Lambda)
\prod_{i\neq k}{\cal L}(\data_k| \Lambda)
\pi(\Lambda)
}\,.
\label{eqn_odds_full}
\end{align}
Finally, we can simplify Eq.~\eqref{eqn_odds_full} to
\begin{align}
\odds_{\Hs_i/\Hn_i}(\alldata)=
\frac{
\int d\Lambda\,{\cal L}(\data_i| \Lambda, \Hs_i)
{\cal L}(\Hs_i|\Lambda)
\pi(\Lambda | \alldata_{i\neq k})
}{
\int d\Lambda\,
{\cal L}(\data_i| \Lambda, \Hn_i)
{\cal L}(\Hn_i|\Lambda)
\pi(\Lambda | \alldata_{i\neq k})
}\,,
\label{eqn_odds}
\end{align}
where $\pi(\Lambda | \alldata_{i\neq k})$ is the posterior distribution on the hyperparameters, conditional on the context data. This expression allows calculation of the odds marginalizing over uncertainty about the priors, both at the model-parameters and hypothesis level, using the contextual data. To provide some intuition and demonstrate the power of this general method we now discuss two specific cases.

First, let $P(A_i | \Lambda) \equiv \xi$ be a hyperparameter modelling the
prior probability of model $A$ for a segment. If $\xi$ constitutes the only
hyperparameter, e.g., $P(\data_i | A_i, \Lambda) = P(\data_i | A_i)$ (and
similarly for the denominator), then
\begin{align}
\odds_{A_i / B_i}(\alldata) & =
\frac{{\cal L}(\data_i| A_i)} {{\cal L}(\data_i| B_i)}
\frac{ \int \xi \pi(\xi | \alldata_{i\neq k}) d\xi}
{\int (1-\xi) \pi(\xi | \alldata_{i\neq k}) d\xi}\\
& =
\frac{{\cal L}(\data_i| A_i)} {{\cal L}(\data_i| B_i)}
\frac{E[\xi]}{1-E[\xi]}\,,
\end{align}
where $E[\xi]$ is the expectation value of $\xi$.
The first factor here is the usual Bayes factor in support of signal in the $i$th
segment, while the second factor is the prior-odds conditioned on all other
data. In the limit of small $E[\xi]$, this is approximately $E[\xi]$. This
expression agrees with how the prior odds is typically defined.

Second, consider the case where model $A$ is a
signal in addition to white Gaussian noise with an unknown standard deviation
$\sigma$. Then, we hyperparameterise by allowing $\sigma \sim
\mathrm{Normal}(\mu_\sigma, \sigma_\sigma)$. If the number of segments of data
in $\alldata_{i\ne k}$ is suitably large, then $\sigma$ can be precisely
estimated from the contextual data. In the language of our hyperparameters, this would imply that
$\sigma_\sigma/\mu_\sigma \ll 1$ and the term $\pi(\mu_\sigma, \sigma_\sigma |
\alldata_{i\neq k}) \approx \delta(\sigma - \mu_\sigma)$. The consequence would
be that $\odds_{\Hs_i/\Hn_i}(\alldata) \approx \odds_{\Hs_i/\Hn_i}(\data_i)$
with a prior $\sigma = \mu_\sigma$ (i.e., the precisely measured value taken
from all the other data).

\section{Recycling inference}
\label{sec:recycling}
In the case of a model $\hyp$ with model parameters $\theta$, for some choice of hyperparameterws $\Lambda$, the hypothesis-evidence is given by
\begin{align}
    {\cal L}(\data | \Lambda, \hyp) = \int d\theta\, {\cal L}(\data | \theta, \Lambda, \hyp) \pi(\theta| \hyp)\,.
\end{align}
Typically, $\theta$ can be of high dimension and this integration itself, done numerically,
can take from a few minutes to many hours depending on the problem in hand. If one then
wants to calculate as part of a posterior inference over $\Lambda$, say, this could
be computationally challenging.

Instead, consider that we compute this marginalization once
at a fixed value of $\Lambda =\Lambda'$ and we have a set of samples $\{\Theta_{i, k}\}$
(where $i$ labels the data segment, and $k$ the sample number) and an
evidence estimate ${\cal L}(\data| \Lambda', \hyp)$. Then,
since $\Lambda$ is a hyperparameter,
${\cal L}(\data| \theta, \Lambda,  \hyp)
={\cal L}(\data| \theta, \Lambda',  \hyp)$
because $\theta$ is fully specified. Therefore
\begin{align}
{\cal L}(\data| \Lambda, \hyp) =
\int d\theta\,
{\cal L}(\data| \theta, \Lambda', \hyp)
\pi(\theta | \Lambda, \hyp)\,.
\end{align}
Applying Bayes rule to expand the likelihood dependent on $\Lambda'$,
noting that ${\cal L}(\Lambda'| \data, \theta) = {\cal L}(\Lambda'| \data)$
and applying Bayes rule again, we find that
\begin{align}
{\cal L}(\data| \Lambda, \hyp_j) =
{\cal L}(\data| \Lambda')
\int d\theta
{\cal L}(\theta| \data) \frac{\pi(\theta| \Lambda)}{\pi(\theta| \Lambda')}\,.
\end{align}
The first term here is the evidence at the fixed value of $\Lambda'$ that we have
already computed. The second term, given the samples generated in computing the evidence
can be approximated by an average over the $k$ samples
\begin{align}
P(\data| \Lambda, \hyp_j) =
P(\data| \Lambda')
\left\langle
\frac{P(\{\Theta_{i, k}\}| \Lambda)}{P(\{\Theta_{i, k}\}| \Lambda')}
\right\rangle_k\,.
\label{eqn_hyperpe}
\end{align}
This allows for rapid evaluation of the likelihood, without having to performed calculations of the high-dimensional nested integrals. One alternative to this method is to apply a kernel-density estimation method, see, e.g. \citet{pitkin2018}. 

We note that, while the above discussion was given assuming that the samples
where computed at a fixed value of $\Lambda$, the same general principle applies given any prior choice (even one that did not arise from the same family) \citep{thrane2019}.
\end{document}